\documentclass[12pt]{article}
\usepackage{graphicx}
\usepackage{color}

\def\Ca{{\rm Ca}^{2+}}
\def\Cacon{[{\rm Ca}^{2+}]}
\def\Caconin{[{\rm Ca}^{2+}]_{\rm i}}
\def\Caconout{[{\rm Ca}^{2+}]_{\rm o}}

\def\Na{{\rm Na}^{+}}
\def\Nacon{[{\rm Na}^{+}]}
\def\Naconin{[{\rm Na}^{+}]_{\rm i}}
\def\Naconout{[{\rm Na}^{+}]_{\rm o}}

\def\Nacons{[{\rm Na}^{+}]_{\rm ns}}

\def\ncxers{$\Na/\Ca$ exchangers}

\def\trp{TRPC6}

\def\sm{smooth muscle}
\def\smc{smooth muscle cell}
\def\smcs{smooth muscle cells}

\def\nm{{\rm nm}}
\def\mum{\mu{\rm m}}

\def\mus{\mu{\rm s}}

\def\beq{\begin{equation}}
\def\endeq{\end{equation}}
\def\bt{\begin{table}}
\def\endt{\end{table}}
\def\bfig{\begin{figure}}
\def\endfig{\end{figure}}

\long\def\symbolfootnote[#1]#2{\begingroup%
\def\thefootnote{\fnsymbol{footnote}}\footnote[#1]{#2}\endgroup}

\begin{document}
\bibliographystyle{unsrt}
\thispagestyle{empty}
\baselineskip 5 mm
\begin{center}
{\bf {\large 
A model for the generation of localized transient $\Nacon$ elevations in 
vascular \sm}}
\vspace{3mm}

\noindent Nicola Fameli,\symbolfootnote[2]{Corresponding author. Address:  
2176, Health Sciences Mall, Vancouver, B. C., Canada V6T 1Z3\\
tel 1-604-8226198; fax 1-604-8226012.
email: {\tt fameli@interchange.ubc.ca}}Kuo-Hsing Kuo, Cornelis van Breemen\\
{\it Department of Anesthesiology, Pharmacology and Therapeutics} \\ 
{\it The University of British Columbia}\\
\end{center}
\vspace{5mm}

{\centerline{\large Abstract}}
\vspace{3mm}

We present a stochastic computational model
 to study the
mechanism of signalling between a source
and a target ionic transporter, both
localized on the plasma membrane (PM)
and in intracellular nanometre-scale subplasmalemmal
signalling compartments comprising the PM, the sarcoplasmic reticulum
(SR), $\Ca$ and $\Na$ transporters,
and the intervening cytosol. We refer to these compartments, sometimes
called junctions, as
cytoplasmic nanospaces or nanodomains.
In the chain of events leading to $\Ca$ influx for SR
reloading during asynchronous
$\Ca$ waves in vascular smooth muscle (VSM),
the physical and functional link between
non-selective cation channels (NSCC)
and $\Na$/$\Ca$ exchangers (NCX)
needs to be elucidated in view of two recent findings: the
identification of the transient receptor potential canonical channel 6
(TRPC6)
as a crucial NSCC in VSM cells and the
observation of localized cytosolic $\Nacon$ transients following purinergic
stimulation of these cells. Having previously
 helped clarify the $\Ca$ signalling step
between NCX and SERCA behind SR $\Ca$
refilling, this quantitative approach
now allows us to
model the upstream linkage of NSCC and NCX. We have implemented
a random walk (RW) Monte Carlo (MC) model with simulations mimicking
a $\Na$ diffusion process originating at the NSCC within PM-SR
junctions.
 The model
calculates the average $\Nacon$ in the nanospace and also
produces $\Nacon$ profiles as a function of distance from the
$\Na$ source.
Our results highlight the necessity of a strategic
juxtaposition of the relevant signalling channels as well as
other physical structures within the nanospaces to
permit adequate $\Nacon$ build-up to provoke NCX reversal and
$\Ca$ influx to refill the SR.
\vspace{2mm}

\pagebreak

\section{Introduction}
We are not used to thinking of ionic sodium ($\Na$)
 as a \textit{signalling} ion, 
despite its known relevance for vascular disease. On the
other hand, the importance of the second messenger $\Ca$ in
signalling cell function is undisputed. There is however recent
experimental 
evidence supporting the idea 
that this important species has, at least in vascular \smcs\
(VSMC),
an almost equally important signalling partner in 
$\Na$\cite{Poburko2007,Poburko2008},
particularly in signalling events that are allowed by the
juxtaposition of signalplexes and transporter carrying membranes at
nanometric distances from one another.
This observation, and others outlined below, 
as well as a wealth of accumulated knowledge on $\Na$ and
$\Ca$ transporters in VSMC has prompted us to take a more quantitative
look at the question of $\Na$-related signalling in PM-SR cytoplasmic
nanospaces (also referred to as nanodomains or junctions):
 nanometre-scale 
signalling compartments comprising the PM, the sarcoplasmic reticulum
(SR), $\Ca$, $\Na$ and other ionic 
transporters (channels, exchangers and pumps) and
relevant signalplexes therein, and the intervening cytosol.

In this article we present a quantitative model 
aimed at elucidating the
mechanism of selective (or site-specific) signalling between a source
ionic transporter and a target ionic transporter, both of which are
localized on the same membrane and are part of a nanodomain.
We developed this model using 
 a stochastic method based on the
simulation of ionic diffusion by random walks (RW) within nanospaces
modeled according to experimental observations.  
Here we concentrate on the
specific example of a $\Na$ transporter, typically a NSCC
and a NCX as its target. 
Generally, during these events $\Na$ entry via a NSCC generates a large 
$\Nacon$ gradient, which, in turn, enables reversal of a NCX
(presumably in the vicinity of the NSCC) and consequent NCX-mediated
$\Ca$ entry into the subplasmalemmal nanodomain. 
A few examples highlighting the biological importance of 
this intra-membrane system 
and its link to pathogenic mechanisms are (table \ref{NCX_patho_list}
summarizes these cases): \textbf{1)}  A critical
subplasmalemmal step in the $\Ca$ signalling cascade giving rise to VSM
cell contraction following G-protein coupled receptor adrenergic
stimulation\cite{Lee2001}---where
 blocking NCX reversal causes attenuation and elimination
of $\Ca$ oscillations due to impairment of SR refilling; replacement
of $\Caconin$ oscillations with a tonic $\Ca$ signal causes a dramatic
decrease in \sm\ force development\cite{Syyong2009};
in pulmonary artery SM, upregulation of the NCX and $\Ca$ entry via
reverse NCX action is considered one of the mechanisms behind elevated
$\Cacon$ in idiopathic pulmonary arterial hypertension 
patients\cite{Zhang2007b};
\textbf{2)} 
Receptor activated $\Na$ entry may also
 induce NCX reversal in endothelial cells (EC) and
this, in turn, gives rise to selective $\Ca$-stimulated eNOS activity and NO
production; eNOS derived NO is an important physiological vasodilator
agent and is accepted as an independent marker of vascular 
health\cite{Teubl1999,Szewczyk2007}; in EC NCX operating
in forward mode is also important in regulating both $\Caconin$ and
$\Cacon_{\rm SR}$; this suggests that local $\Naconin$,
besides the cell membrane potential and the equilibrium potential of the
NCX,
have a role in the regulation of forward NCX 
too\cite{Nazer1998a,Blaustein1999};
\textbf{3)} In nerve terminals, a transient $\Nacon$ increase linked to a
tetanic pulse of the action potential can reverse the NCX to induce
presynaptic potentiation; this observation implies
 a role for the NCX in synaptic
facilitation and has consequences for short
term memory from reverse NCX malfunction\cite{Zhong2001,Poburko2008};
 \textbf{4)} In skeletal
muscle NCX plays an important role in $\Ca$ homeostasis,
operating in forward as well as reverse mode in that function;
$\Na$/$\Ca$ exchange is involved in the control of muscle fatigue 
and there are
reports  supporting the notion that 
the beneficial role of external $\Ca$ in protecting
slow-contracting soleus muscle against high-frequency fatigue
 depends mostly on $\Ca$ influx through
reversal of the NCX\cite{Germinario2008}; \textbf{5)} although the role of
NCX in heart 
is still poorly understood, both its $\Ca$-efflux and influx modes are
observed in cardiac myocytes and and the latter is likely a consequence of
subplasmalemmal $\Na$ elevations\cite{Levesque1991}.

All of the above emphasize that modulation of $\Ca$ signaling via $\Na$
and $\Na$/$\Ca$ exchange is of great clinical relevance in areas such
as hypertension, chronic heart failure and possibly
cerebral and skeletal muscle malfunction.

This study sheds new light on 
a few new scientifically interesting key factors regarding the
``workings'' of intracellular signalling
nanospaces. We investigate the role, and importance, of
having a confining surface, namely another lipid membrane, facing the
membrane where the source and target transporters belong. This emerges
as an important feature of nanodomains, as previous observations had
hinted\cite{Lee2005a,Fameli2007},
but it would appear that it alone cannot be entirely responsible for
the generation of a sufficiently large local $\Nacon$ transient elevations.  
We find that to understand how
sufficient $\Nacon$ can be built up within the nanodomain to trigger
NCX reversal and
$\Ca$ signalling downstream, it is important to
account for other factors such as the role of physical obstructions to $\Na$ 
diffusion and the possible organization of these obstructions.
\vspace{5mm}

\begin{table}
\begin{center}
    \begin{tabular}{rlll} \hline\hline
	System & NCX mode/function & Function & References\\\hline
	VSM & rev/SR refilling & blood flow & \cite{Lee2001}\\
	    & during ACaW      &              & \\
        PASM & rev/NCX upregulation & blood flow & \cite{Zhang2007b}\\
        EC & rev and fwd/eNOS activity & NO
production regulation & \cite{Teubl1999}\\
        nerve & rev/tetanic pulse & short memory & 
		\cite{Zhong2001,Poburko2008}\\
	      & of action potential & & \\
        skeletal muscle & rev/$\Ca$ influx  & muscle fatigue &
\cite{Germinario2008}\\
        \hline\hline
    \end{tabular}
    \caption{Sampling of systems in which NCX modes are linked to
pathologies. Legend: rev=reverse mode ($\Ca$ influx); fwd=forward mode 
($\Ca$ efflux); ACaW=Asynchronous $\Ca$ waves; 
PASM=pulmonary artery smooth muscle; EC=endothelial
cells.}\label{NCX_patho_list}
\end{center}    
\end{table}

\section{Methods}
\subsection{Electron microscopy}\label{TEMmethods}
Details of the electron microscopy have been described
previously\cite{kuo2001}. 
The primary fixative solution contained 1.5\%
glutaraldehyde, 1.5\% paraformaldehyde and 1\% tannic acid in 0.1~M
sodium cacodylate buffer that was pre-warmed to the same temperature
as the experimental buffer solution (37~$^{\circ}$C). The rings of
rabbit IVC were fixed at 37~$^{\circ}$C for 10 minutes, then
dissected into small blocks, approximately
1~mm$\times$0.5~mm$\times$0.2~mm in size and put in the same fixative
for 2 h at 4~$^{\circ}$C on a shaker. The blocks were then washed
three times in 0.1~M sodium cacodylate (30 min). In the process of
secondary fixation, the blocks were post-fixed with 1\% OsO4 in 0.1~M
sodium cacodylate buffer for 1 h followed by three washes with
distilled water (30 min). The blocks were then further treated with
1\% uranyl acetate for 1 h ({\it en bloc} staining) followed by three
washes with distilled water. Increasing concentrations of ethanol
(25, 50, 70, 80, 90 and 95\%) were used (10 min each) in the process
of
gradient dehydration. 100\% ethanol and propylene oxide were used
(three 10 min washes each) for the final process of dehydration. The
blocks were infiltrated in the resin (TAAB 812)
and then embedded in molds and polymerized in an oven at
60~$^{\circ}$C for 8 h. The embedded blocks were sectioned on
a microtome using a diamond knife at the thickness of 80~nm. The
sections were then stained with 1\% uranyl acetate and
Reynolds lead citrate for 4 and 3 min, respectively. Images were
obtained with a Phillips 300 electron microscope at 80~kV.

\subsection{Simulations}
Simulations of $\Na$ diffusive motion from a source transporter are
based on an implementation of a
Monte Carlo random walk (RW), along the lines of methodology
previously employed for $\Ca$ transport simulations\cite{Fameli2007}.
The model nanospace in which the simulations of $\Na$ diffusion take
place arises from observed physical features and properties
of the essential elements for nanodomain signalling. These are
either obtained from our laboratory's studies or from the available
literature\cite{Lee2002b} and they 
are essentially the TRPC6 cytosolic `radius' and $V_{\rm
max}$\cite{Dietrich2007b}, the 
typical dimensions of intracellular nanospace ultrastructure, estimates 
on the number of pillars, $\Na$ diffusivity in cytosol, 
expected $\Nacon$ necessary for NCX reversal (see section
\ref{foundation})
and during localized $\Na$ transients (LNats) observed in \cite{Poburko2007}.  
Fig. \ref{barebox} is a to-scale representation of the geometry of the
model nanospace used in the simulations.
\bfig[tb]\centering
\includegraphics[scale=.4]{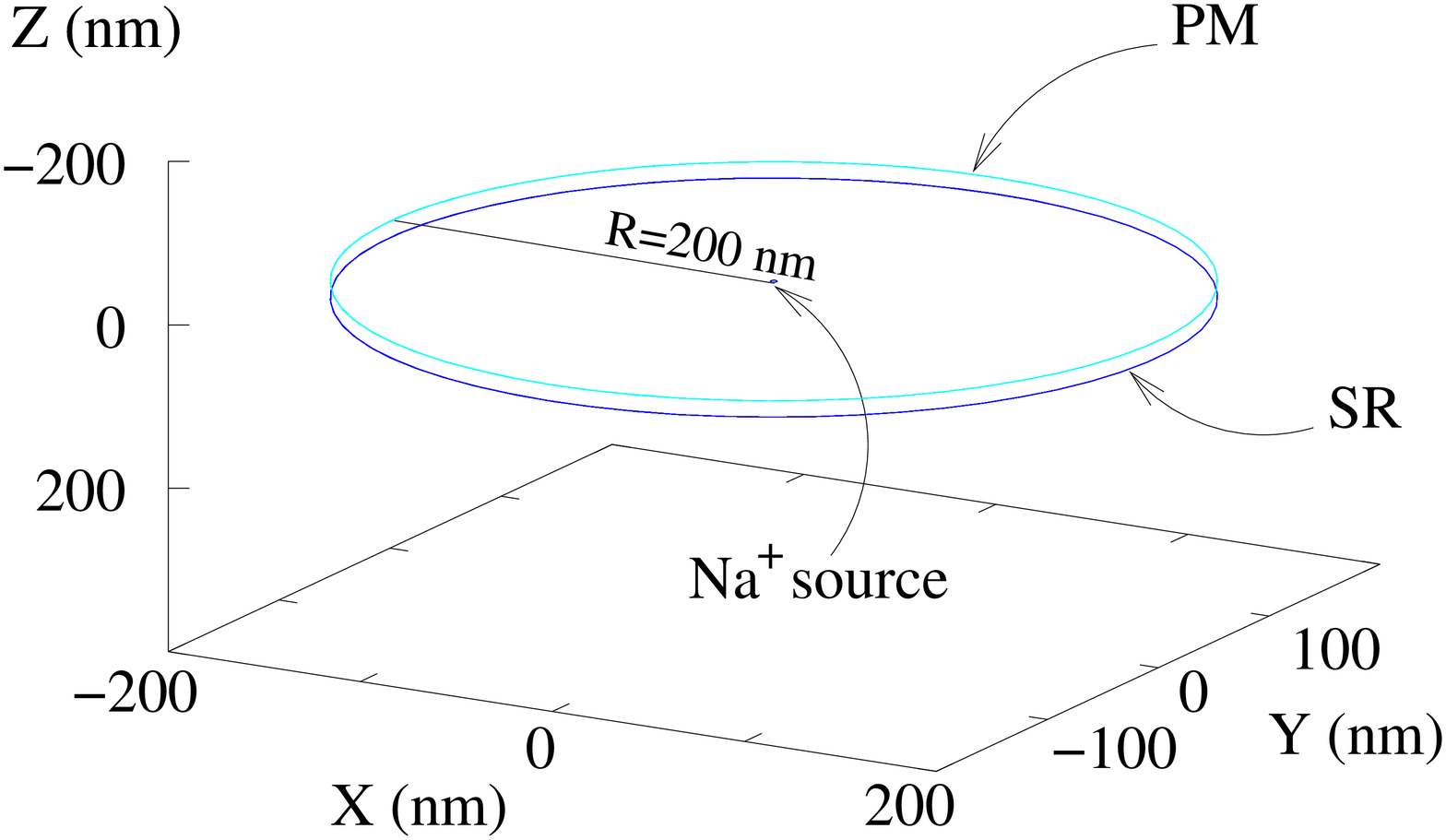}
\caption{To-scale model nanospace used in the simulations. The
separation between the two surfaces (cyan and blue) is 20 nm.}
\label{barebox}
\endfig
In the simulations, particles representing $\Na$ performs a RW 
on a cubic lattice with spacing $s=0.2$ nm; initially picked as an 
approximation
to the expected $\Na$ mean free path in water (in turn, as an estimate to
the mean free path in cytosol), we also carried out tests for effects of
varying this parameter in a previous article 
and revealed no substantial influence on the results\cite{Fameli2007}.
The RW time step $\tau$ was chosen by running several simulations letting
particles cover a predetermined straight line distance $d$, and recording
the number of RW steps $N$ taken to cover $d$. 
From diffusion theory, the total time $t$
taken by a random walker in three dimensions 
to cover the distance $d$ is $t=d^2/(6D_{\rm meas})$, where
$D_{\rm meas}$ is the measured diffusivity of $\Na$ in muscle
cytosol\cite{Kushmerick1969b}. The quantity $t/N$ was our 
choice for $\tau$ and its value is $10^{-11}$ s. 
 
The boundary conditions in our simulations are as follows. 
At the PM and SR membranes we implemented reflecting conditions: 
ions arriving at one of those surfaces during
their RW are reflected back into the nanospace.
At the edges of the model nanospace conditions are perfectly
absorbing: ions reaching the lateral boundary of the junction are
considered absorbed by the cytosol external to the nanospace, 
lost from the junctional population and no longer contributing
to the PM-SR nanospace concentration, $\Nacons$.
As explained later, we also positioned a number of obstacles (we refer
to them as pillars) to $\Na$ motion 
spanning the distance between the membrane in the nanospace. 
 Pillars behave like elastic scatterers for $\Na$.

Simulation code is written and tested in the C programming language
 on a computer running a Linux operating system. 
After testing and troubleshooting, programs are recompiled and
run in one of the WestGrid computing nodes\cite{wg}.
The pseudo-random number generator we used is the algorithm
\texttt{gsl\_rng\_m19937} of the GNU Scientific Library\cite{gsl,gsl_URL}, 
since it has sufficient randomness
and quality requirements for our purposes.
The plots with
simulation results were produced using the ``freely distributed plotting
utility'' \texttt{gnuplot}\cite{gnuplot}.

\section{Results}
\subsection{Model foundation}\label{foundation}
This laboratory has previously experimentally
established that a $\Nacon$ transient from a NSCC enabled the generation
within a PM-SR junctional nanospace of a sufficiently high $\Nacon$
gradient to cause NCX reversal\cite{Lee2001}. 
A quantitative estimate of the level of
such burst can be obtained by comparing the equilibrium potential of
the NCX, $E_{\rm NCX}$, with the membrane potential, $V_{\rm M}$, using
typical values for vascular \smcs. These quantities are linked by the 
equation $E_{\rm NCX}=V_{\rm M}$ (where $E_{\rm NCX}=
3E_{\rm Na}-2E_{\rm Ca}$)
between the electrochemical potential of $\Na$ ($E_{\rm Na}$),
of $\Ca$ ($E_{\rm Ca}$) and the membrane potential ($V_{\rm
M}$)\cite{Blaustein1999}. 
(In general, $E_{\rm
C^{z+}}=RT/({\rm z}F)\,{\rm ln}([C^{\rm z+}]_{\rm out}/[C^{\rm
z+}]_{\rm in})$, where $R$ is the universal gas constant, $T$ the
absolute temperature, ${\rm C}^{\rm z+}$ is a $z$-valent cation, and
$F$ is Faraday's constant.) 
NCX reversal occurs when $E_{\rm NCX}< V_{\rm
M}$. We can study this inequality for both resting and activated cell
conditions by a plot like the one in Fig. \ref{Encx-Vm}. 
Using for an activated cell $V_{\rm M}=-20$ mV,
$\Naconout=140$ mM, $\Caconin=10\mum$, $\Caconout=2$ mM,
$R=8.3$~J/(mol K), $T=310$~K, and $F=9.65\times 10^4$~J/(V mol),
we observe that, during activation, 
a $\Nacon$ transient of the order of 30 mM or
greater is necessary to cause NCX reversal. 
In this exercise, we have used
an estimated value for $\Caconin=10\mum$, as that is approximately 
the lowest value $\Caconin$ we expect from observations cited 
earlier\cite{Fameli2007}.
\bfig[tb]\centering
\includegraphics[scale=.4, angle=-90]{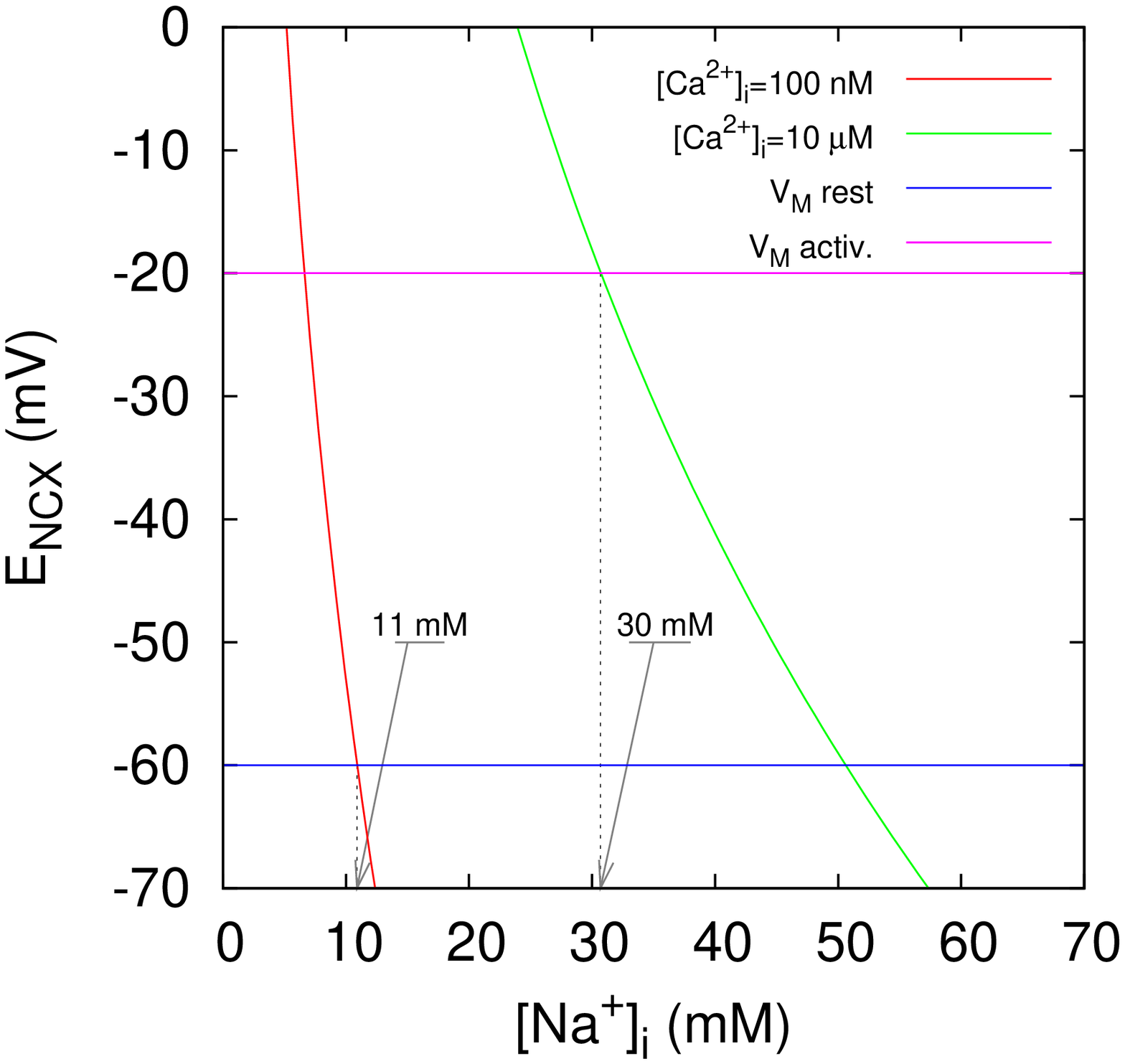}
\caption{Equilibrium NCX potential in the case of a resting (red curve)
or activated (green) \smc\ compared with the corresponding membrane
potential, V$_{\rm M}$.}
\label{Encx-Vm}
\endfig

Now, equipped with \textbf{(a)} the 
fundamental observation of localized $\Nacon$ elevation transients in full 
agreement with the values suggested by the study of Fig.
\ref{Encx-Vm}\cite{Poburko2007}, 
\textbf{(b)} better
knowledge of
 the identity of NSCC as TRPC6\cite{Poburko2007,Lemos2007},   
\textbf{(c)} the basic idea
that the presence of intracellular nanospaces 
is necessary for this signalplex to be
complete, we propose a model to investigate the role of strategic placement of
transporters with respect to each other, as well as of a 
confining membrane and other ionic diffusion limiting structures 
in the generation within these nanospaces of sufficiently high
$\Nacons$ to permit NCX reversal.

The model nanospace used for the study is illustrated in Fig.
\ref{barebox}. The dimensions expressed therein are based on high
quality EM images showing that the PM-SR separation in these nanospaces
is remarkably uniform and about 20 nm. Lateral extension of these
closely apposed PM-SR regions is approximately 400 nm (see
\cite{Fameli2007}).

\subsection{Random walk simulations: bare PM-SR nanospace}
 The simplest model nanodomain we studied consists 
of a shallow cylinder-shaped volume of height $h=20\;\nm$ and
radius $R=200\;\nm$, as in Fig. \ref{barebox}. $\Na$ entering the
nanodomain via an NSCC are represented
as particles doing a RW based on a given diffusivity $D=600\mum^2/s$,
corresponding to that of $\Na$ in muscle cytoplasm as reported in
\cite{Kushmerick1969b}.
For simplicity, there is only one $\Na$ source
positioned at the centre of the PM side of the nanospace (in section
\ref{discussion} we will
elaborate further on the issue of the number of $\Na$ sources).
The simulation programs output the computed $\Nacons$ rise above
resting level as a function of
time, thereby giving an average $\Nacon$ increase in the nanospace.
(From now on $\Nacons$ denotes the increase in cytoplasmic $\Na$
concentration in the nanodomain between PM and SR.)
Results from this set of simulations are 
shown in Fig. \ref{Na_vs_tr_bare}, left 
panel. 
\bfig[tb]\centering
\includegraphics[scale=.55, angle=-90]{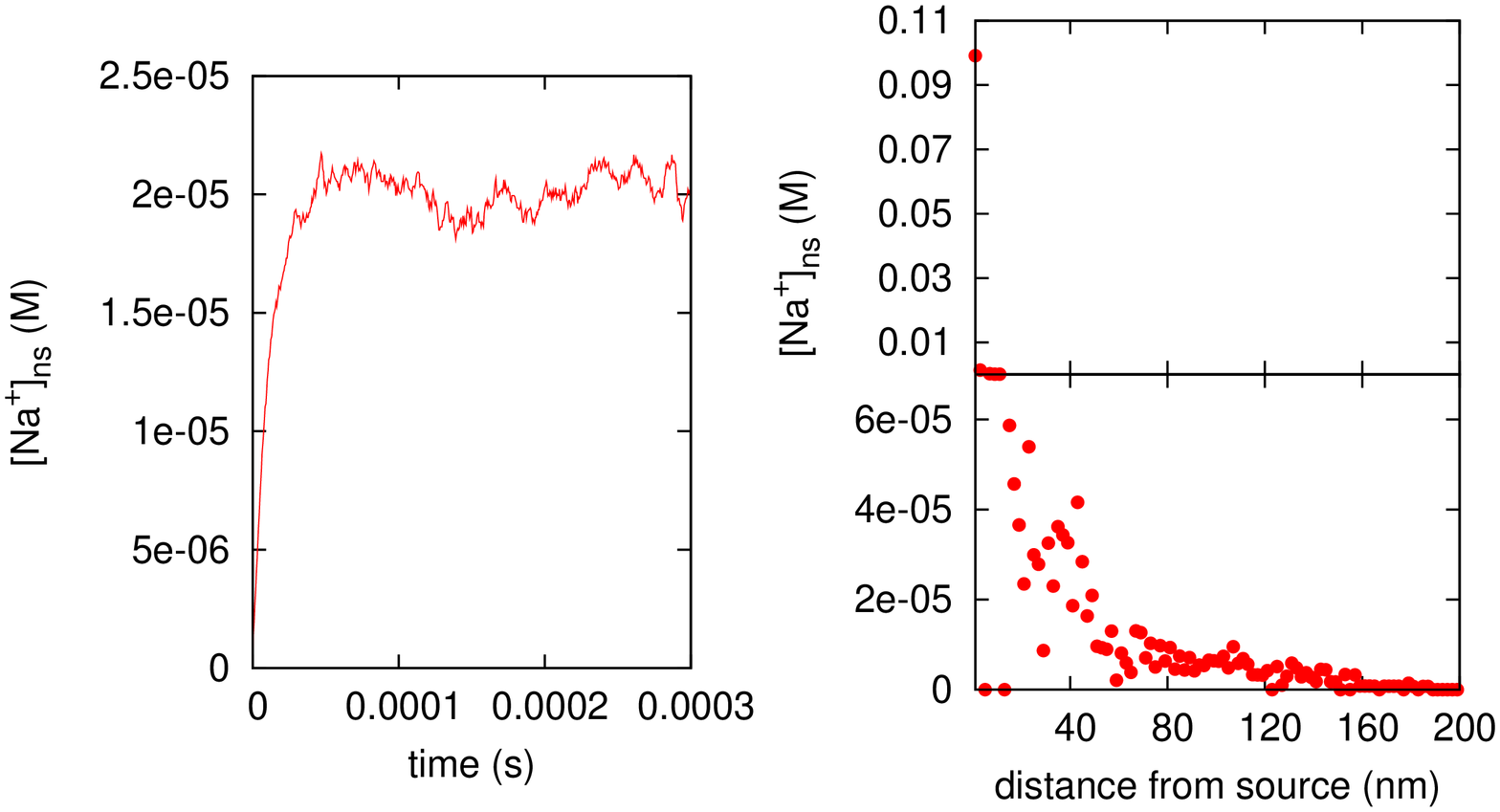}
\caption{Left panel: $\Nacons$ vs t for simplest form of nanospace as
in Fig. \ref{barebox}; right panel:  $\Nacons$ vs distance from $\Na$
source placed at the centre of the PM side of the nanospace.
}\label{Na_vs_tr_bare}
\endfig
This plot helps
establish the time scale after which we can consider that the $\Nacon$ has
reached a steady-state level. We can observe from the graph that after
approximately 100 $\mus$ the concentration has reached a plateau after
having increased from zero during an initial transient, as we have explained in
the previous section.
Having established a time scale for the formation of the approximately
maximum level of $\Nacons$, we compute and study the
concentration profile inside the junction, by plotting $\Nacons$ as a
function of the distance from the $\Na$ source. 
The graph in the right panel of 
Fig. \ref{Na_vs_tr_bare} illustrates a representative
result.

\subsection{Random walk simulations: randomly distributed obstacles in
the nanospace}
Evidently, this
simplest incarnation of the model is inadequate to describe the
generation of $\Na$ transients of the observed size\cite{Poburko2007}, 
since at steady state elevation of
$\Nacons$ hovers around $2\times10^{-5}$ M or about three orders of
magnitude less than the observed values of 15--20 mM. 
 We need to consider other
nanospace features emerging from our ultrastructural images that may be
responsible for a larger increase in the $\Nacons$. Barring
artificially changing
the value of $\Na$ diffusivity, $\Nacons$ can be ``forced'' to increase if
the ions were able to dwell longer in the nanospace than they are in
the simple version of the system analyzed so far.


There is convincing evidence suggesting the existence of structures
spanning the width of the nanospace and which could constitute an
impediment to the free diffusion of $\Na$\cite{Poburko2008,Devine1972}. 
Our own observations confirm the
existence of 
electron opaque ``pillars'' in transmission
electron microscopy images like the one in Fig. \ref{pillars}. The size and 
abundance of these electron opaque structures compares well with the 
electron dense ``bridges'' observed by Devine and collaborators in the early 
`70s\cite{Devine1972}.
\bfig[tb]\centering
\includegraphics[scale=.4]{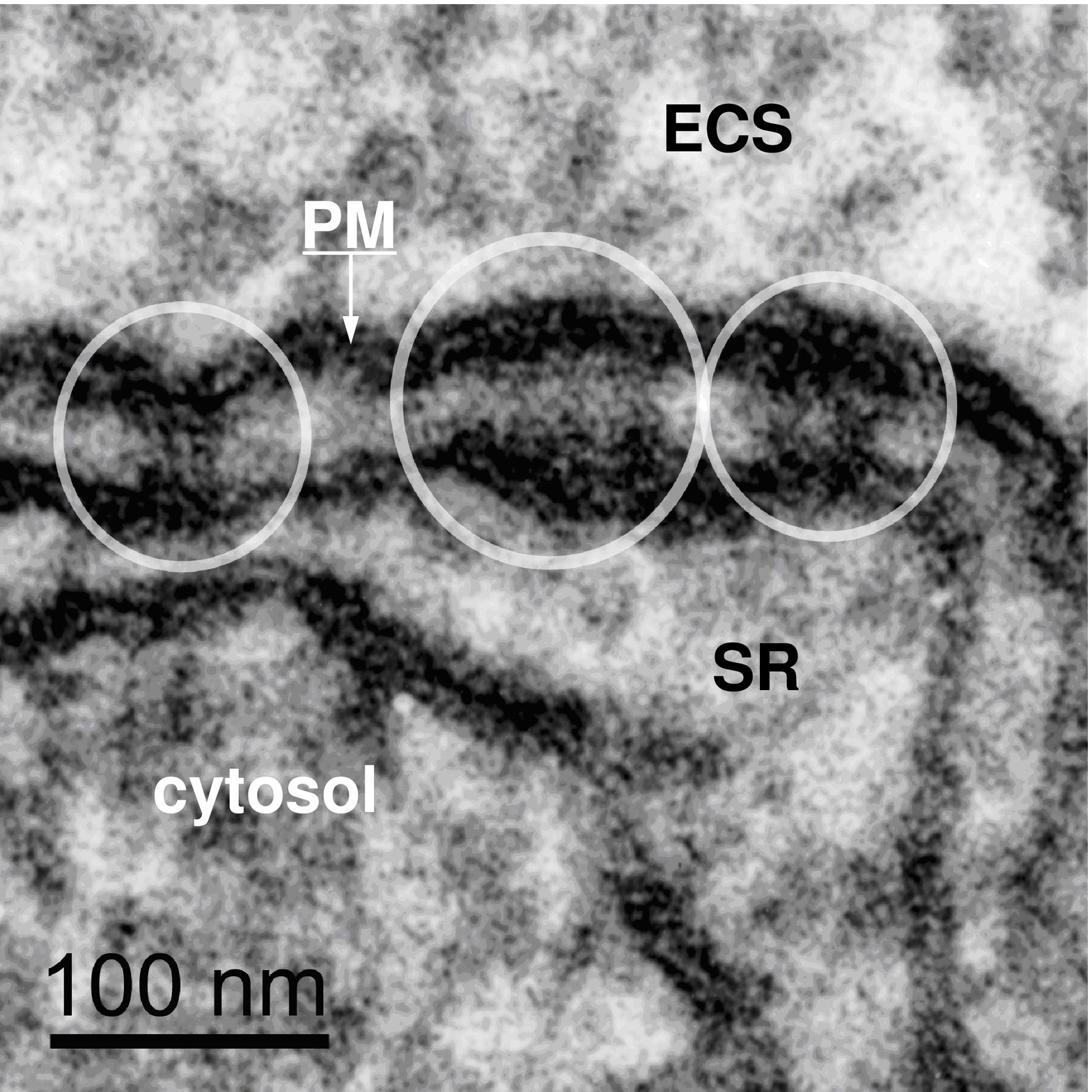}
\caption{Transmission electron microscopy image showing nanospace
bridging electron opaque structures (white circles). Tissue: rabbit inferior
vena cava. SR=sarcoplasmic reticulum; PM=plasma membrane;
ECS=extracellular space.}
\label{pillars}
\endfig

Keeping the overall geometry of the model nanospace the same (Fig. 
\ref{barebox}), we have therefore implemented a number of junction
spanning structures in the form of cylindrical pillars, having
estimated their size from several images like the one in Fig.
\ref{pillars}. From those same images it is possible to approximate the
percentage junctional volume occupied by those structures and, in turn,
an approximate number of them expected per nanospace. In a series of
simulations, we have
represented up to 200 pillars randomly distributed within the
nanospace (Fig. \ref{200_p_box}), 
and then simulated the $\Na$ diffusion by a random walk
within the pillar-populated nanospace. 
\bfig[tb]\centering
\includegraphics[scale=.4, angle=-90]{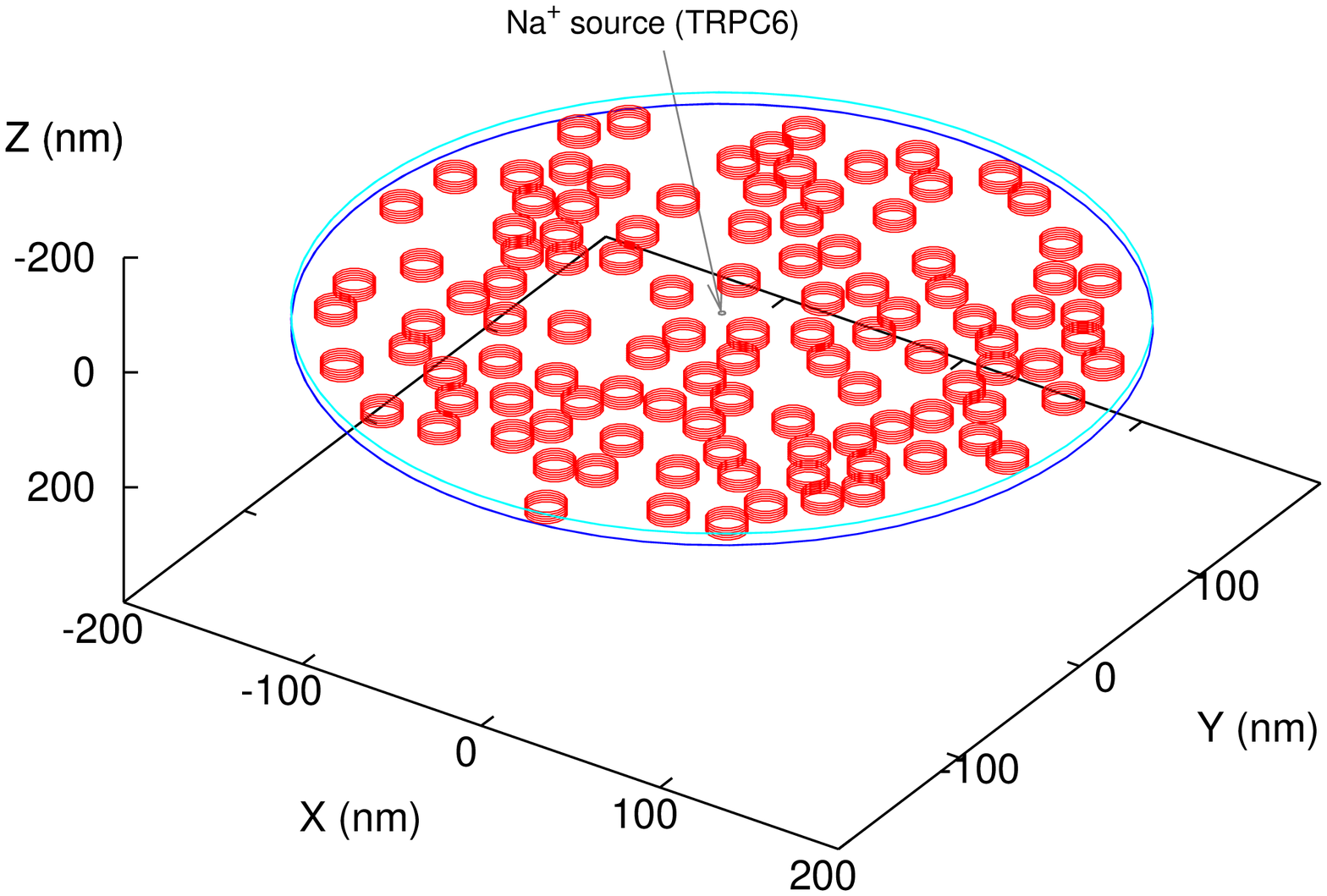}
\caption{To-scale model nanospace including about 200 10-nm radius
randomly placed
cylindrical pillars spanning the distance between the two surfaces.}
\label{200_p_box}
\endfig
In this case too, 
we let the simulations run for a
time sufficiently long to ensure that a steady-state level for the
$\Nacons$ was established. 
Results are reported in Fig. \ref{Na_vs_tr_pil}.
In all simulations involving random positioning of pillars, 
to minimize bias from the particular random pillar distribution, 
we have take average values of the
computed $\Nacon$ over 10 different random pillar distributions. These
are the values plotted in the graphs we present in this article.
\bfig[tb]\centering
\includegraphics[scale=.55, angle=-90]{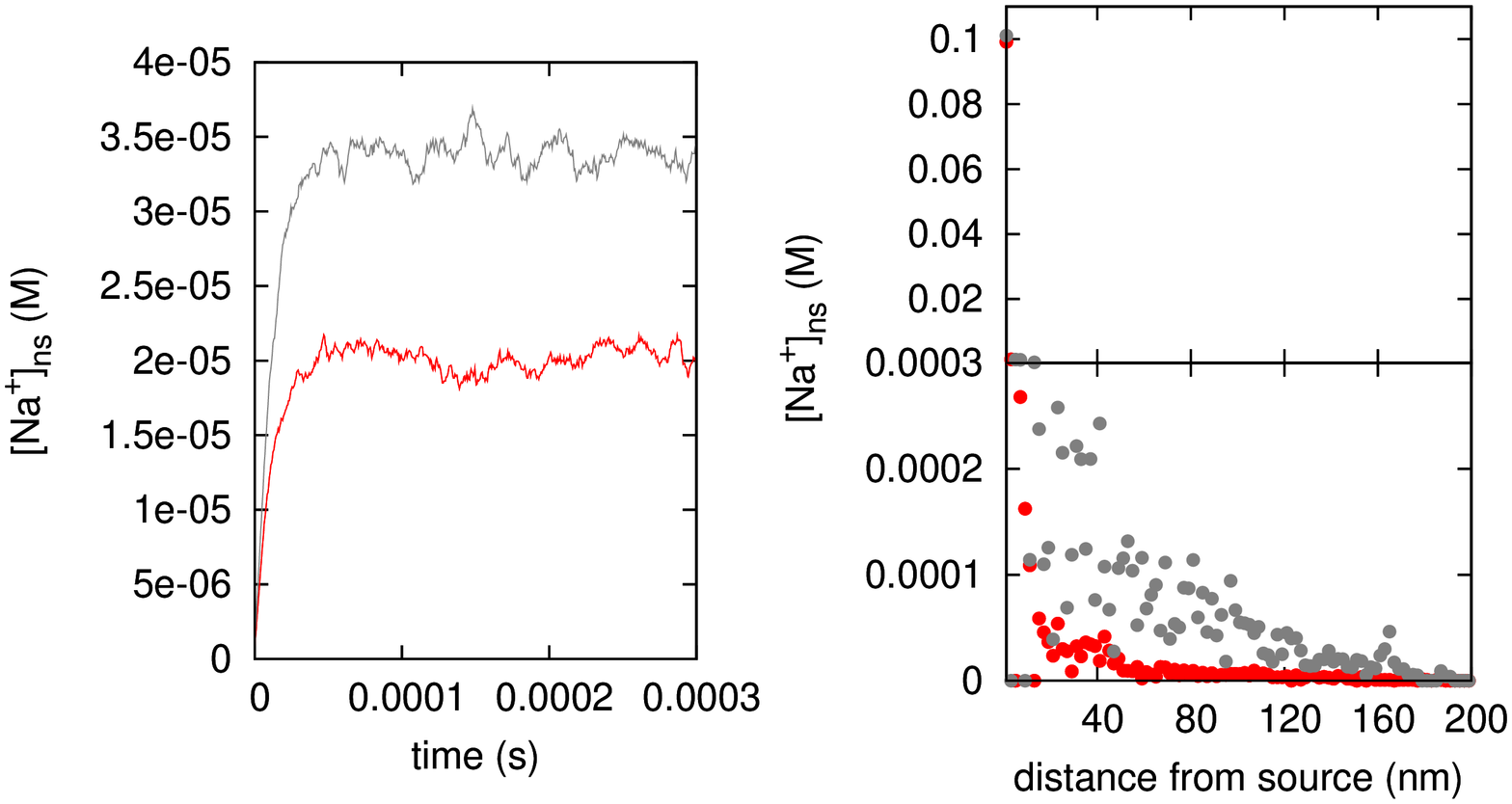}
\caption{$\Nacons$ within a model nanospace containing a number of
pillars occupying about 30\% of the volume, as in Fig. \ref{200_p_box}. 
Left panel: computed $\Nacons$ as a function of time; the values of
$\Nacon$ in this graph are an average across the entire nanospace.
Right panel: concentration profile within nanospace. The red curve and
dots are the same data displayed in Fig. \ref{Na_vs_tr_bare}.}
\label{Na_vs_tr_pil}
\endfig

The results of this series of simulations 
indicate that having some form of
impediment to ionic motion in the junction does produce the effect of
increasing the $\Nacons$ and of changing its profile to one that decays 
more slowly with distance from the $\Na$ source. We
ensured that this effect was not 
simply a consequence of the decreased nanospace
volume due to the presence of the pillars by plotting the steady-state
$\Nacon$ computed with different numbers of pillars in the junction
(blue dots in Fig. \ref{pillar_effect}) and comparing it with an
increase in $\Nacon$ merely due to reducing the nanospace volume by the
volume of the pillars (red line in Fig. \ref{pillar_effect}). 
The plot in Fig.
\ref{pillar_effect} demonstrates that ion collisions with 
pillars do indeed have a
role in forcing $\Na$ to dwell longer in the nanospace.
\bfig[tb]\centering
\includegraphics[scale=.4, angle=-90]{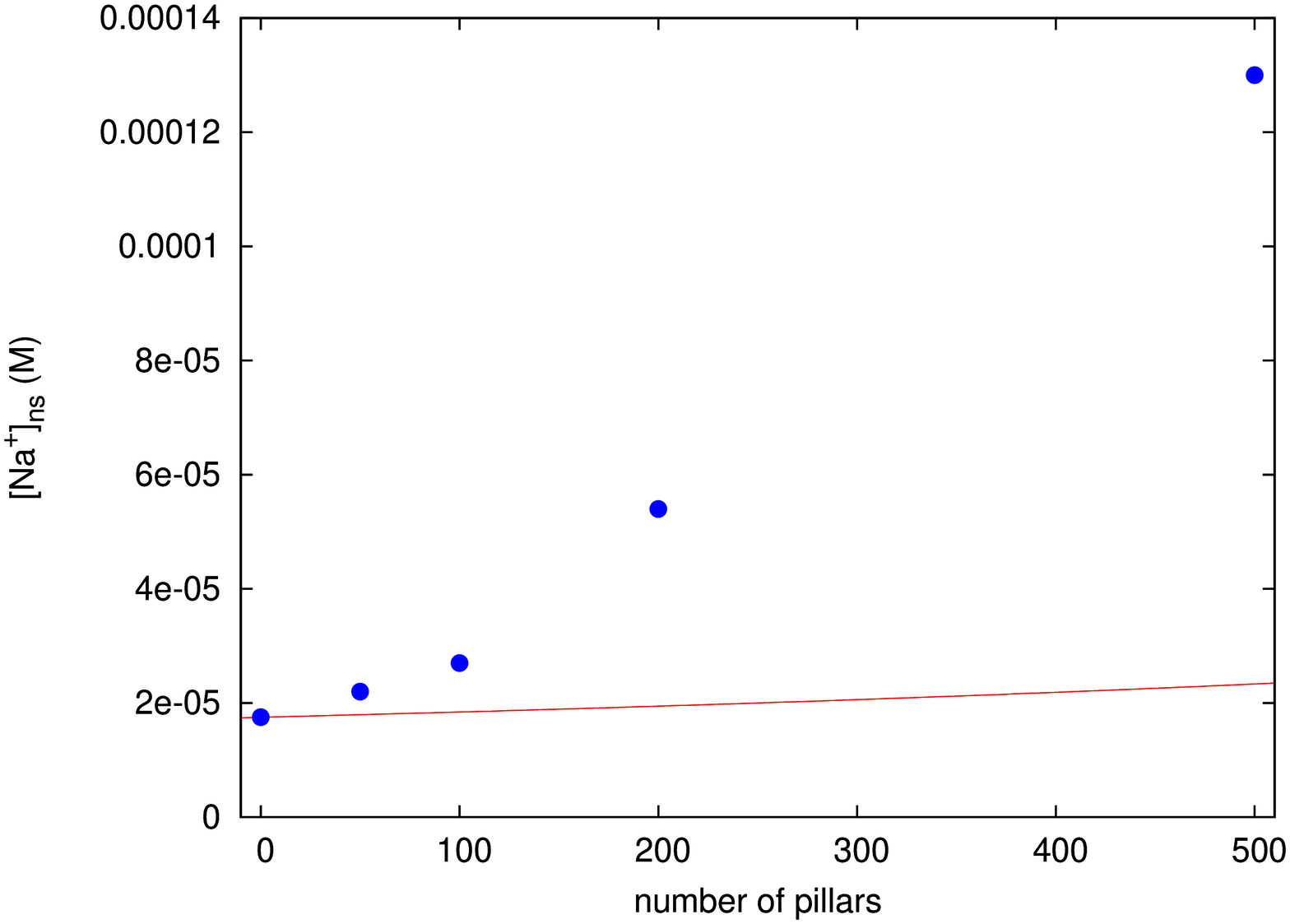}
\caption{Comparison of steady-state $\Na$ calculated only accounting
for the volume effect of pillars in the nanospace (red line) and
computed from the simulations (blue dots).}
\label{pillar_effect}
\endfig

\subsection{Random walk simulations: non-randomly distributed obstacles}
Clearly, the presence of obstacles to diffusion has an effect of
increasing $\Nacons$, however the values we obtain this way are not yet
comparable with those measured during the local $\Nacon$ elevation 
transients. Other
junctional features need to be accounted for in order to understand the
mechanism giving rise to such high $\Na$ transients necessary to
reverse the NCX and observed
by Poburko and coworkers\cite{Poburko2007}. 
We considered the hypothesis that nature might place these obstacles 
``strategically'' rather than randomly, in a neighbourhood of a $\Na$
source, so as to favour the generation of the gradients needed to drive
the signalling chain. The rationale is that while a random set of
pillars does show an ability to retain ions in the nanospace longer and
therefore allow higher concentration build up, it does not do it
efficiently enough to quantitatively account for the observed $\Nacons$
transients. We then ran some simulations in which pillars are placed in
a circle around a $\Na$ source in such a way that we can control the
porosity of this pillar fence to the passage of random walking ions.
(Imagining to stand where the $\Na$ source is, the circle of pillars
would appear as a 20-nm-high set of slabs surrounding the source
itself, with thin rectangular gaps between the slabs; with
this in mind, porosity is defined as the ratio of the entire surface
area of the gaps
to that of the slabs plus gaps.) 
The configuration of our model nanospace in this case is shown
in Fig. \ref{circle_14}, as a two dimensional $x$,$y$-projection.
\bfig[tb]\centering
\includegraphics[scale=.45, angle=-90]{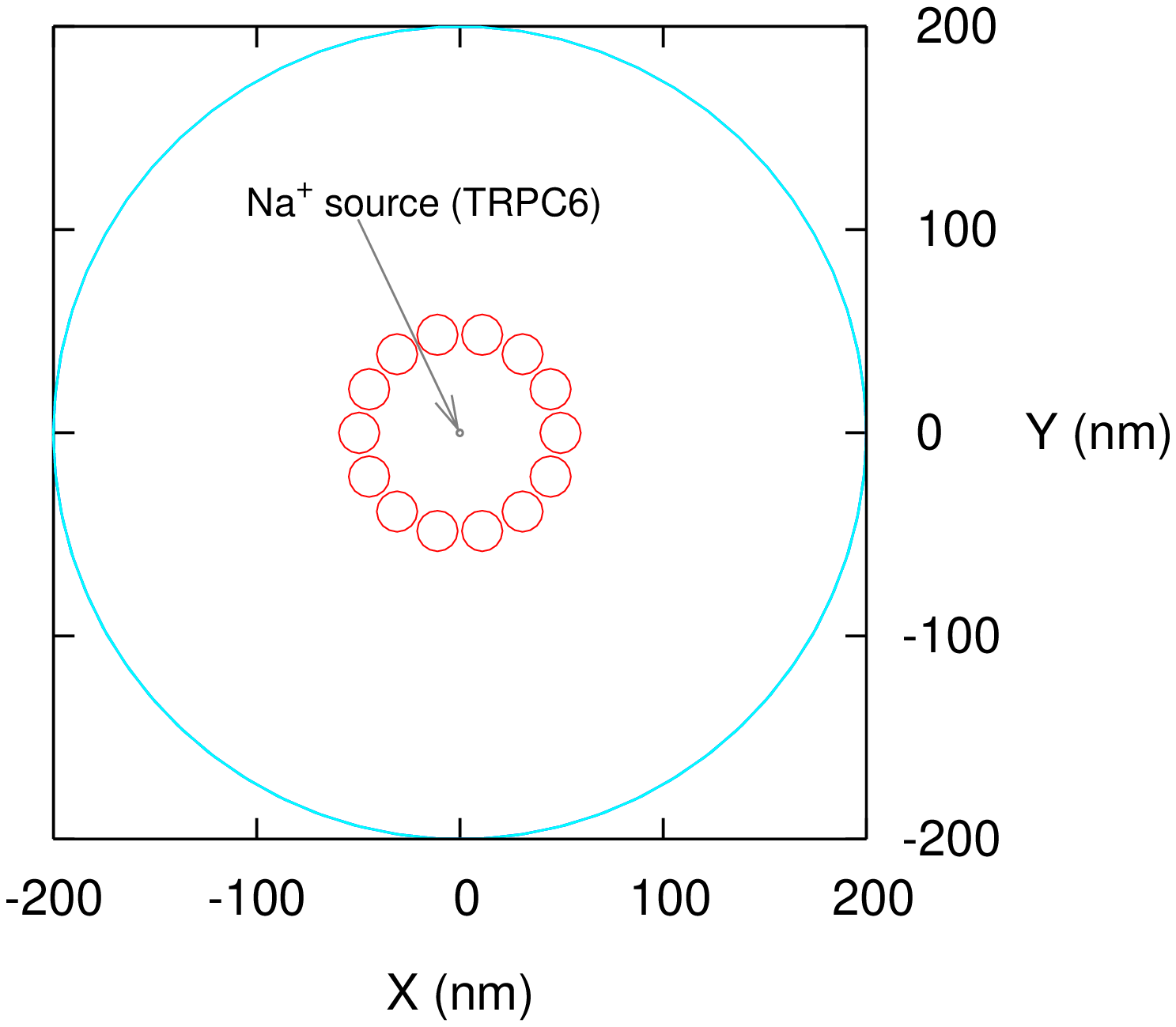}
\caption{To-scale $x$,$y$-projection of model nanospace including 
14 10-nm-radius
cylindrical pillars positioned in a 5-nm-radius 
ring around a $\Na$ source in the
centre, thereby giving a 1\% porosity for ion passage.}
\label{circle_14}
\endfig

Representative results from these simulations are reported in Fig.
\ref{Na_vs_tr_cop}.
\bfig[tb]\centering
\includegraphics[scale=.55, angle=-90]{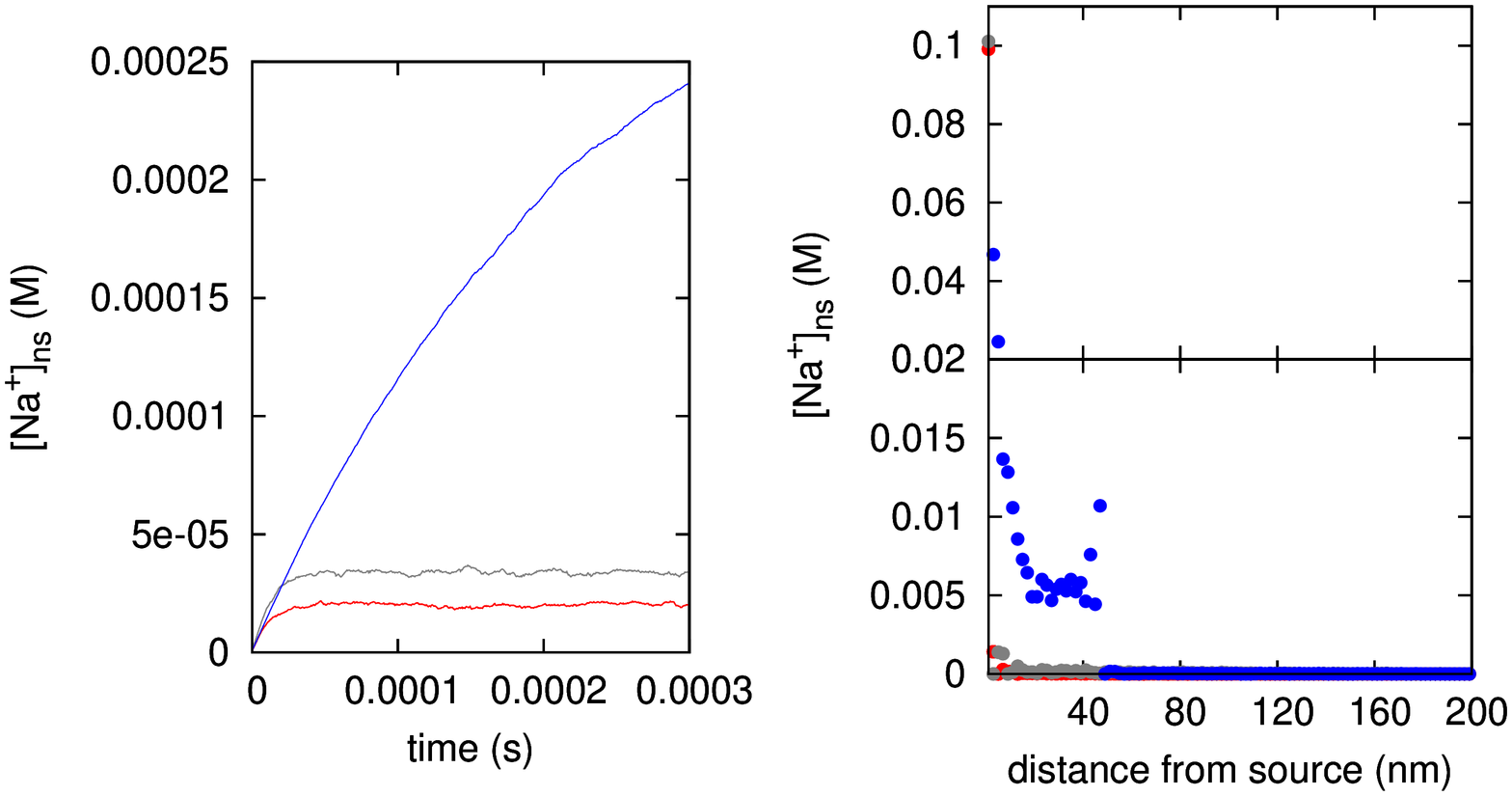}
\caption{
$\Nacons$ within a model nanospace containing 14
pillars ``strategically'' positioned around a $\Na$ source at the
centre and making a 1\% porous fence, as in Fig \ref{circle_14}.
Left panel: computed $\Nacons$ as a function of time.
Right panel: concentration profile within nanospace at $t=0.0003$ s.
The red and the grey curves and dots refer to data displayed in Fig.
\ref{Na_vs_tr_bare} and \ref{Na_vs_tr_pil}.
}
\label{Na_vs_tr_cop}
\endfig
The dramatic increase in the $\Nacons$ in the vicinity of the source
caused by this type of obstacle configuration is immediately evident.
$\Nacons$ is in this case of the same order of magnitude as the
observed localized
$\Nacon$ transient elevation phenomenon\cite{Poburko2007}. Note
furthermore that in this configuration the model suggests that the time
for the $\Nacons$ to reach steady-state is much longer than in the
case of random distribution of pillars. The more scattered data in the
plot of the right panel in Fig. \ref{Na_vs_tr_cop} are due to the fact
that the system in this case has yet to reach equilibrium.

\section{Discussion}\label{discussion}
The stochastic computational model presented herein attempts to give a
quantitative description of the mechanism behind the observed localized
$\Nacon$ transients observed in VSMC\cite{Poburko2007}. Based as
much as feasible on experimental observations of the physical and
physiological features of the intracellular nanospaces in which these
transients
are hypothesized to occur, the model results lead us to conclude that
$\Nacon$ can build to sufficiently high values in PM-SR junctions to
give rise to the observed transients. 

Three main steps lead to the fundamental hypothesis
behind this work. The first stems from earlier work by this laboratory
elucidating the sequence leading to asynchronous $\Ca$ oscillations in
VSMC\cite{Poburko2007,Fameli2007}. Succinctly, a large external $\Na$ influx causes the reversal of
the NCX and consequent $\Ca$ entry to refill the SR after SR-$\Ca$
release via IP$_3$R channels upon cell stimulation provokes contraction.
The second is the observation of two main features of the localized
$\Nacon$ elevation transients\cite{Poburko2007}: 
they appear as a punctate pattern on the periphery of VSMC 
(with puncta having a given time course), and their peak $\Nacon$
values are comparable to the $\Naconin$ values necessary to cause NCX
reversal (see Fig. \ref{Encx-Vm} and relevant text). 
Thirdly, our previously published model
supports the idea that NCX reversal-mediated $\Ca$ entry in PM-SR nanospaces
introduces sufficient $\Ca$ to refill the SR during asynchronous $\Ca$
waves\cite{Fameli2007}. 
Based on these three points, the hypothesis we set out to study with
our model is that the observed
localized $\Nacon$ transients occur in PM-SR nanodomains (or
nanospaces or junctions), in other words, that in those junctions, due to $\Na$
influx via a TRPC6 channel, $\Nacon$ can attain levels, that
cause $\Ca$ entry via NCX reversal.

Simulation results from our simplest version of model 
(Fig. \ref{Na_vs_tr_bare}), namely,
a PM-SR nanospace filled with cytosol
(represented in our simulations by the diffusivity of $\Na$) 
with only one TRPC6 channel as 
$\Na$ source suggest that this simple view is not
adequate to describe the formation of $\Nacon$ transients. This is
mainly due to the large value of the diffusion coefficient of $\Na$,
which is about three times that of free $\Ca$ in cytosol, and the fact that 
there is no observed buffering effect of $\Na$ in 
cytosol\cite{Kushmerick1969b}.
Recent $\Nacon$ measurements in isolated
rabbit ventricular myocytes by
Despa and Bers (\cite{Despa2003a,Despa2004})
also suggest that the effective diffusivity of $\Na$ may be much slower
than the one found by Kushmerick and Podolsky\cite{Kushmerick1969b}, 
although no explanation
as to the mechanism behind it was suggested. In \cite{Despa2003a},
among other things, Despa and Bers measured endogenous $\Na$ buffering
in
cytosol and found it negligible
when compared to that of other
important species like $\Ca$, for example. It is well known that $\Ca$
buffering lowers its diffusivity dramatically\cite{allbritton1992,wagner1994}, 
thus possibly contributing to easier generation of
$\Ca$
gradients in confined spaces like the PM-SR nanospaces.
Presuming the slight $\Na$ buffering
effect observed in cardiac cell cytosol is mirrored in \smcs, it
clearly could not provide sufficient slowing down of $\Na$
diffusivity to aid local transient creation. It seems instead
ever more plausible that it must be some sort of physical obstruction
to ionic motion that gives rise to a slower effective diffusivity for
$\Na$.
We then considered that physical obstructions to ionic motion can have
the overall effect of increasing the time spent by each $\Na$ in the
nanospace thereby increasing the value of $\Nacons$. 
 Ever since first reported measurement of $\Na$ diffusivity in muscle
cells\cite{Kushmerick1969b}, it was hinted
that it is physical rather than chemical
interactions that cause a retardation of the ionic (and non-ionic for
that matter---sorbitol and sucrose specifically) intracellular
diffusion.
Our ultrastructural observations of nanospace spanning electron opaque 
structures support this idea (Fig. \ref{pillars}) 
and agree fully with earlier reports of
similar structures in SMC\cite{Devine1972}.
The introduction of such physical obstruction in the simulations seems then 
more than warranted and very plausible. Moreover, if the ``tortuosity
factor'', introduced by Kushmerick,
arising from physical
interaction is indeed important in the bulk cytosol, then it would be 
even more so in
the restricted PM-SR nanospaces. 
While the introduction of physical obstacles to ionic motion in our
simulations does produce higher $\Nacons$ gains (Fig. \ref{Na_vs_tr_pil}
and \ref{pillar_effect}), to obtain a quantitative agreement with the
observed level of $\Nacon$ during the transients we need to hypothesize
further that these junction spanning structures are not distributed at
random, but rather form an organized
barrier around a given $\Na$ source within the nanospace. Our
simulation results in this case (Fig. \ref{Na_vs_tr_cop}) agree
quantitatively with the observed $\Nacon$ transients, giving a $\Na$ of
the same order of magnitude as the experimental measurements.
The left panel plot in Fig. \ref{Na_vs_tr_cop} also shows that the
$\Nacon$ with this configuration of pillars will reach steady state with a time constant that
is at least ten times longer than in the case of randomly distributed
pillars. This is another feature supporting our hypothesis that these
transients are formed in the PM-SR nanospaces with the aid of a relatively 
tight
fence of pillars around the $\Na$ source. In fact, the measured
transients are reported to be very long lived with a ramp up time to
steady state of about 30 s, a time scale that the simpler versions of
the model could not reproduce at all.

It therefore appears that two features are essential if these transients must
occur within PM-SR nanospaces: there must be physical obstructions to
$\Na$ motion, forming an organized barrier around each $\Na$ source
(TRPC6 channels) {\em and\/} NCX must be localized near a TRPC6 within
such barrier to be able to sense the high $\Nacon$, reverse 
and allow $\Ca$ entry. The second of these two features is
suggested by a number of observations.  There are studies supporting
the idea that TRPC channels and \ncxers\ are
functionally and physically linked and form an important signalplex in
several different systems\cite{Eder2005}.
Still other studies indicate
that at least certain kinds of TRP channels are found in close
proximity of caveolin\cite{Lockwich2001,Brazer2003}. Combined with our
earlier observations that NCX tend to crowd near the necks of 
caveolae\cite{Fameli2007},
this makes a strong case for a physical association between TRP
channels and NCX. 
 The first of the above mentioned desirable features---patterns of 
physical obstructions within the nanodomains---is
a harder matter to study experimentally and one of the current/future
directions our laboratory is following.

To further study the issue of the localized $\Nacon$ transient 
duration requires more
computationally demanding simulations, which we are currently tackling.
We can however make a qualitative argument to suggest how the time
scale of the observed transients can also be supported by our model. 
We have so far only implemented one $\Na$ source, or one TRPC6 channel. 
However, we have observed earlier that
during activation of rabbit inferior vena cava SMC there could be about 16 NCX
in the junctions\cite{Fameli2007}. If we conjecture that TRPC6 and NCX
must be in physical proximity, as we speculated above, 
it is consistent to assume 
that there may be more than one \trp\ in each junction
functioning as a $\Na$ entry gate, and perhaps even as many as there
are NCX. 

Other transporters such as NCX and $\Na$/K$^+$ ATPases (NKA) certainly
also play
a role in shaping the PM-SR nanodomain $\Nacon$ profiles that drive
NCX-mediated $\Ca$ entry during VSMC activation. In the simple model we
presented in this article, we focussed on the role of TRPC6 since it is
by far the larger capacity transporter of the TRPC6, NCX, NKA trio
(millions of ions per second, hundreds per second and tens per second,
respectively). This first approximations quantitative model supports
the idea that localized $\Nacon$ elevation transients take place at
PM-SR nanodomains and strongly suggests that physical impediment to
ionic mobility of $\Na$ is also a necessary factor for their
generation.

\section*{Acknowledgments}

The research was
supported by grants from the
Canadian Institute of Health Research and the Heart and Stroke
Foundation of British Columbia and Yukon.
This project has been enabled by the use of WestGrid computing resources, 
which are funded in part by the Canada Foundation for Innovation, 
Alberta Innovation and Science, BC Advanced Education, and the 
participating research institutions. WestGrid equipment is provided by IBM, 
Hewlett Packard and SGI.


\bibliography{in-mem-sig}
\end{document}